\documentclass[aps,prd,twocolumn, groupedaddress,preprintnumbers,nofootinbib]{revtex4-1}
\usepackage{graphicx,amssymb,amsmath,color,slashed}
\usepackage{enumerate}
\usepackage{hyperref}

\def\beq{\begin{equation}}
\def\eeq{\end{equation}}

\def\bea{\begin{eqnarray}}
\def\eea{\end{eqnarray}}
\def\bei{\begin{itemize}}
\def\eei{\end{itemize}}
\def\bmat{\begin{matrix}}
\def\emat{\end{matrix}}
\def\ble{\begin{flushleft}}
\def\ele{\end{flushleft}}
\def\={\,=\,}
\def\+{\,+\,}
\def\-{\,-\,}

\def\GeV{\,{\rm GeV}\,}
\def\TeV{\,{\rm TeV}\,}

\def\TeV{\,{\rm TeV}}
\def\GeV{\,{\rm GeV}}

\def\AFB{A_{\rm FB}}

\newcommand{\Fig}[1]{Fig.~\ref{#1}}
\newcommand{\Eq}[1]{Eq.(\ref{#1})}

\begin{document}

\title{Correlation of top asymmetries: loop versus tree origins}

\author{Sunghoon Jung}
\email{nejsh21@gmail.com}

\author{P. Ko}

\author{Yeo Woong Yoon}
\email{ywyoon@kias.re.kr}

\author{Chaehyun Yu}

\affiliation{\vspace{1mm} School of Physics, Korea Institute for Advanced Study, Seoul 130-722, Korea}

\begin{abstract}
We study the correlation of top asymmetries that are sensitive to the different origin of (a new contribution to) the total asymmetry: loop- or tree-level  origins. We find that both the size and sign of the correlation between total and $t\bar{t}j$ inclusive asymmetries are inherently different depending on the origin. We demonstrate the correlation by using the color-singlet $Z^\prime$ and the pure axigluon taken as representative models of loop- and tree-induced total asymmetries. We calculate the next-to-leading order QCD corrections to the $Z^\prime$ and perform Monte-Carlo event generation. The correlation is understood in the QCD eikonal approximation using its color structure. 
\end{abstract}

\preprint{KIAS-P14028}

\maketitle

\section{Introduction}

The top forward-backward asymmetry, $A_{\rm FB}$, provides valuable information of the underlying production mechanisms of top pairs as well as higher-order QCD. The standard model(SM) $\AFB$ arises first at next-to-leading order(NLO) QCD top pair production which had been 
estimated in Refs.~\cite{Kuhn:1998jr,Kuhn:1998kw} based on earlier works on total cross-section~\cite{Nason:1987xz,Beenakker:1988bq,Nason:1989zy,Beenakker:1990maa,Mangano:1991jk} and asymmetry~\cite{Berends:1973fd,Halzen:1987xd}. Since the Tevatron measurements, the prediction is refined by performing resummations~\cite{Almeida:2008ug,Ahrens:2011uf} and electroweak corrections~\cite{Bernreuther:2005is,Kuhn:2005it,Manohar:2012rs}, and QED effects on $\AFB$ were especially shown to be positive and dominant among them~\cite{Hollik:2011ps,Bernreuther:2012sx,Kuhn:2011ri}. 

The inclusive $A_{\rm FB}$ in the $t\bar{t}j$ sample measures the real corrections to top pair production and is inherently related with the NLO nature of the total asymmetry in the SM. Its Tevatron measurements~\cite{Abazov:2011rq,Aaltonen:2012it} (followed by LHC's~\cite{Aad:2013cea,CMS:2013nfa}) immediately triggered exciting developments of QCD-related subjects: parton showering~\cite{Skands:2012mm,Hoeche:2013mua}, small-$q_T$ resummation technique~\cite{Zhu:2012ts,Li:2013mia} and better calculations of the process~\cite{Dittmaier:2007wz,Dittmaier:2008uj,Melnikov:2010iu,Melnikov:2011qx}. The color structure of (new) production mechanisms may also be measured in this channel~\cite{Gripaios:2013rda}.

Such intimate connection with higher-order nature and the consequent characteristic spectrum of $d A_{\rm FB}/ d p_T(t\bar{t})$ predicted by NLO QCD may make any tree-level new physics contributions more easily measurable and better characterizable in the $t\bar{t}j$ channel against QCD and possible loop-induced new asymmetries. To address the question of how well we can do so, we study correlations of top asymmetries measured in various channels including $t\bar{t}j$ by comparing two new physics models generating top asymmetries \emph{first} at the loop- and tree-level.

There have been many efforts to build and test tree-level $A_{\rm FB}$ models~\cite{Jung:2009jz}. Currently, however, every new large asymmetries are somewhat constrained~\cite{Gresham:2011pa} and no compelling reasons for considering only tree-models are present. Loop-induced \emph{leading} asymmetries are definitely worth studying. Although the loop-induced contribution to the total $A_{\rm FB}$ will be typically small, it is not obvious how large and how well measurable it is in the $t\bar{t}j$ channel, for example. Of course, any model will affect some asymmetry at some higher-order, but if we restrict to the one-loop as an interesting leading order for $A_{\rm FB}$, the model building option for the loop-induced asymmetry is reasonably limited and different from that of the tree-level asymmetry. Furthermore, the consideration may help to better understand QCD and to learn what various $A_{\rm FB}$ can tell us about new physics.

\section{Models and constraints} \label{sec:model}

\begin{figure*}[t]
\includegraphics[width=2\columnwidth]{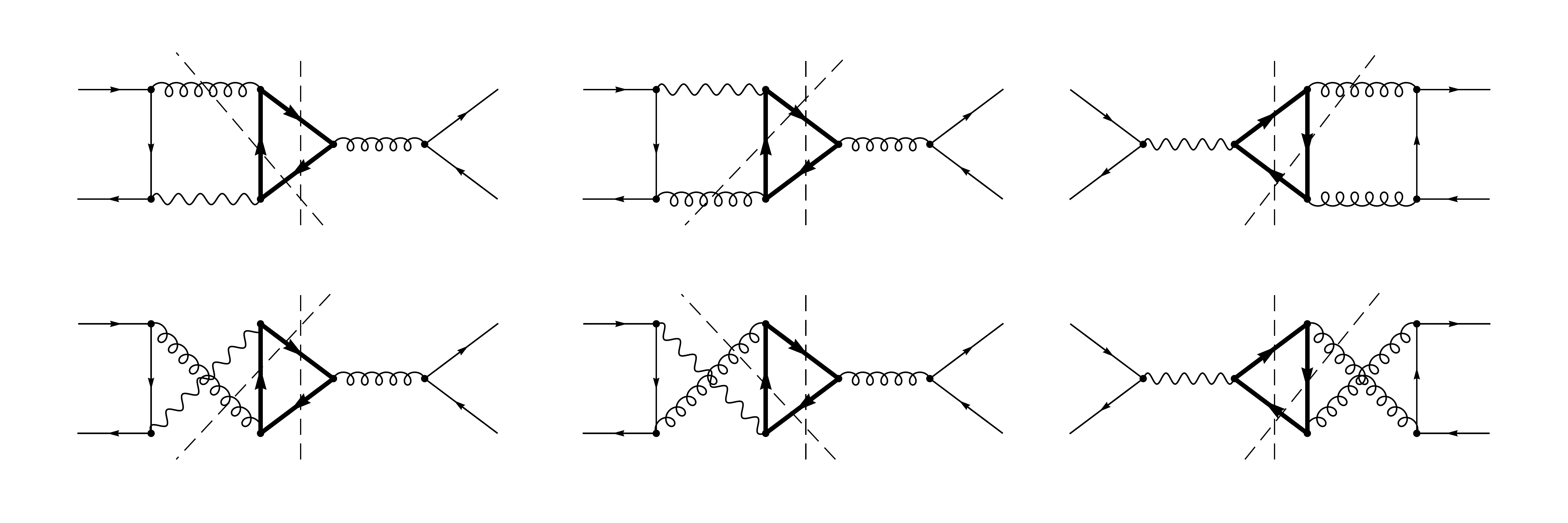}
\caption{Feynman diagrams for the NLO corrections of interference between QCD and the X. Wavy(thick) lines denote heavy X(top) propagators. The dashed lines represent possible cut lines. These are leading contributors to top asymmetry from X.}
\label{fig:nlo} \end{figure*}

It has been categorized~\cite{Jung-RG}, based on the operator mixing analysis, that the only four-quark operator capable of generating sizable top asymmetries \emph{first} at one-loop order without modifying the total cross-section much is $VV^{(1)}$, where $VV^{(1)}$ implies the color-singlet vectorial-vectorial current-current interaction. Meanwhile, it is also known that the $AA^{(8)}$ operator, color-octet axial-axial, induces the asymmetry via the tree-level interference with QCD~\cite{Jung-RG,Jung:2009pi,Degrande:2010kt}. 

Motivated from these studies, we consider the heavy leptophobic $Z^\prime$ as our representative model for the loop-induced asymmetry (``loop-model"). The $Z^\prime$ is color-singlet, spin-1 and denoted by X. It couples equally to left- and right-handed quarks with coupling constant $g_X$
\beq
g_X \, \left( \, \sum_{i=1}^5 \overline{q}_i \gamma^\mu q_i \+ \eta_t \overline{t} \gamma^\mu t  \, \right) \, X_\mu.
\eeq
The coupling to the top quark can have a relatively different sign $\eta_t = \pm1$. We do not refer to any specific models and take the couplings and the mass of $X$ as free parameters.

The axigluon~\cite{oai:arXiv.org:0911.2955} represents the model for the tree-level $A_{\rm FB}$ (``tree-model"). It is color-octet and spin-1. It has purely axial vector couplings and is denoted by AxA
\beq
g_A \, \left( \, \sum_{i=1}^5 \overline{q}_i \gamma^\mu \gamma_5 T^a q_i 
\+ \eta_t \overline{t} \gamma^\mu \gamma_5 T^a t  \, \right) \, A^a_\mu.
\eeq
Again, the couplings and the mass are free parameters.

We assume that both X and AxA are heavy and broad, i.e. $\Gamma / M \sim 0.4$, to avoid dijet resonance searches~\cite{lightaxi1}. Dijet angular distribution $\chi$ searches~\cite{CMS:contact2012,ATLAS:contact2012} are simulated for both models using \texttt{MadGraph}~\cite{Alwall:2011uj}. Conservatively assuming that data agree with SM backgrounds, we find that 3TeV resonances with $g_{X,A} \lesssim 3.5$ are allowed within current experimental uncertainties\footnote{Note that the officially reported bound around 10TeV~\cite{CMS:contact2012,ATLAS:contact2012} is resulted in from the deficit of data compared to SM backgrounds, but we conservatively assume that this is a downward fluctuation.}. Top pair resonance searches~\cite{TheATLAScollaboration:2013dja,Chatrchyan:2013lca} are weaker than dijet resonance searches.

\section{Next-to-leading order corrections}

We carry out the NLO calculation of the X model as well as SM top pair production. The X model's generation of the top asymmetry at ${\cal O}(\alpha_s^2 \alpha_X)$ is exactly analogous to that of QED at ${\cal O}(\alpha_s^2 \alpha_e)$ -- we discuss an important difference below. The X model interferes with QCD via diagrams in \Fig{fig:nlo} to induce the asymmetry. 

The leading QCD and QED contributions at ${\cal O}(\alpha_s^3,\alpha_s^2\alpha_e)$ are also calculated independently in this work. We confirm that our SM results are well consistent with those from \texttt{MCFM}~\cite{Campbell:2000bg} and several previous SM NLO calculations ~\cite{Hollik:2011ps,Bernreuther:2012sx}.

The calculation is performed with dimensional regularization for regularizing ultraviolet(UV) divergences in $d=4-2\varepsilon$. We use $\overline{\rm MS}$ scheme for the renormalization.  We use Feynman gauge for any gauge bosons; we note that the new Goldstone bosons do not couple to SM fermions. For virtual corrections, we carry out Dirac algebra and the reduction to Passarino-Veltman functions using \texttt{FeynCalc}~\cite{Mertig:1990an} in $d$ dimension. The numerical computation for the resulting functions is performed by \texttt{QCDLoop}~\cite{Ellis:2007qk}. We also cross checked our results by independent calculation by using our own in-house Mathematica code based on the Laporta's algorithm~\cite{Laporta:2001dd} for the reduction and state-of-art method for calculating master integrals.

In order to obtain differential cross sections, we write our own Monte-Carlo(MC) event generator. We use the Catani-Seymour's dipole subtraction method~\cite{Catani:1996vz,Catani:2002hc} to systematically handle infrared(IR) divergences for each event points. QED and X dipole functions are identical to QCD ones with proper change of color factors. We cross checked our (integrated) dipoles with \texttt{MadDipole} package~\cite{Frederix:2008hu,Frederix:2010cj}. Vegas integration~\cite{Lepage:1980dq} is adapted for MC phase space integration and event generation. 

As for the SM parameters, we set $m_t = 173.34\GeV$~\cite{ATLAS:2014wva}, $\alpha_e = 1/128.0$ fixed. We employ CTEQ6.6M~\cite{oai:arXiv.org:0802.0007} PDF set for ${\cal O}(\alpha^3)$-contributions and CTEQ6L for ${\cal O}(\alpha^2)$-contributions. The QCD coupling constant at $m_Z$ scale is chosen to be $\alpha_s(m_Z) = 0.118$ conforming with PDF sets. We conveniently choose $\alpha_{CS}=0.1$~\cite{Nagy:1998bb,Nagy:2003tz,Campbell:2004ch} for the X and $\alpha_{CS} =0.2$ for the SM, but we checked that our numerical result is independent on the choice of $\alpha_{CS}$ and that the $\alpha_{CS}$-dependence of individual virtual and real corrections are consistent with \texttt{MCFM} results.

\section{Leading asymmetry contributions}

We use the rest-frame asymmetry defined in terms of rapidity difference
\beq
A_{\rm FB} = 
\frac{ \sigma_{FB} }{ \sigma_{tot} } = 
\frac{ N( \Delta y(t)  >0) - N( \Delta y(t)  <0 ) }{ N( \Delta y(t)  >0) + N( \Delta y(t)  <0 ) },
\eeq
where
\beq
\Delta y(t) = y(t) - y(\bar{t}).
\eeq
The lab-frame asymmetry will be correlated in a mostly model independent way.

Throughout this paper, we consider various top asymmetries: total, $0j$ exclusive and $1j+$ inclusive. The total asymmetry is what is typically measured and mentioned one -- it is measured with $t\bar{t}$+anything sample. The $t\bar{t}$+anything sample is divided into the $t\bar{t}$+0$j$ and $t\bar{t}$+1$j$+anything where the extra jet is conveniently defined to have $p_T \geq 20$GeV. The divided samples correspond to the $0j$ exclusive and $1j+$ inclusive samples. For QCD, the division more or less measures the virtual and real corrections. Note that both $0j$ and $1j+$ are IR-finite individually. 

We use our own NLO MC event generator for the SM and X model predictions while we use \texttt{MadGraph}~\cite{Alwall:2011uj} for AxA models; note that hard radiation can be reliably simulated from \texttt{MadGraph} without showering. The showering would mainly affect the $0j$ sample but only subdominantly as tree-level effects are dominant there.

\vspace{0.1in}

\begin{table}[t] \centering
\begin{tabular}{c|c|| c| c}
\hline \hline
Model & parameters & $\Delta \sigma_{tot}$ & $\Delta \sigma_{FB}$ \\
\hline \hline
X+ & $M_X=3\TeV,\, g_X = 2.0,\, \eta_t = -1$ &  $ 0.11$pb & 55fb \\
X-- & $M_X=3\TeV,\, g_X = 2.51,\, \eta_t = +1$ &  $0.34$pb & -85fb \\
\hline
AxA+ & $M_A=3\TeV,\, g_A = 1.5,\, \eta_t = -1$ &  $\sim 0$pb & 177fb \\
AxA-- & $M_A=3\TeV,\, g_A = 1.2,\, \eta_t = +1$ &  $\sim 0$pb & -92fb \\
\hline
QCD & -- & 5.56pb & 393fb\\
QED & -- & $\sim 0$pb & 76fb\\
\hline \hline
\end{tabular}
\caption{Benchmark parameters and their predictions for Tevatron. Models are defined in text. Leading contributions are only added for each model as described in text and as defined in regard of \Eq{eq:dafbdmtt1} and \Eq{eq:dafbdmtt2}. For reference, QCD and QED results are also shown. Each model contributions are individually shown, and total results are then the sum of all relevant contributions.}
\label{tab:benchmark} \end{table}

We include only leading contributions to both numerator and denominator when calculating asymmetries. For the case of $A_{\rm FB} (1j+)$ measuring the $1j+$ inclusive asymmetry, the leading effects read
\beq
A_{\rm FB} (1j+) \simeq \frac{\alpha_s^3 N^{(1)}_1 + \alpha_s^2 \alpha_e N^{(2)}_1 + \alpha_s^2 \alpha_X N^{(3)}_1 + \dots }{ \alpha_s^3 \,D^{(1)}_1 + \alpha_s^2 \alpha_e D_1^{(2)} + \alpha_s^2 \alpha_X D^{(3)}_1 + \dots }
\label{eq:dafbdptt} \eeq
for both X and AxA models with the proper coupling constant $\alpha_X$ or $\alpha_A$. Both numerator and denominator start at ${\cal O}(\alpha^3)$ in both models, where $\alpha$ can be any relevant couplings. The first two terms in numerator and denominator are pure QCD and QCD-QED interference effects, and the third term is QCD-new physics interference. In our parameter space, ${\cal O}(\alpha_s \alpha_X^2)$ contributions are subdominant. We denote $D_0^{(i)}, N_0^{(i)}$ as contributions at ${\cal O}(\alpha^2)$ and  $D^{(i)}_1, N^{(i)}_1$ as contributions at ${\cal O}(\alpha^3)$.

On the other hand, for $A_{\rm FB} (tot)$ measuring the total asymmetry, the leading effects read
\beq
A_{\rm FB} (tot) \simeq \frac{\alpha_s^3 \tilde{N}^{(1)}_1 + \alpha_s^2 \alpha_e \tilde{N}^{(2)}_1 + \alpha_s^2 \alpha_X \tilde{N}^{(3)}_1 + \dots}{ \alpha_s^2 \tilde{D}^{(1)}_0 + \alpha_e^2 \tilde{D}^{(2)}_0 + \alpha_X^2 \tilde{D}^{(3)}_0 + \dots }
\label{eq:dafbdmtt1} \eeq
for X model and
\beq
A_{\rm FB} (tot) \simeq \frac{\alpha_s^3 \tilde{N}^{(1)}_1 + \alpha_s^2 \alpha_e \tilde{N}^{(2)}_1 + \alpha_s \alpha_A \tilde{N}^{(1)}_0 + \dots}{ \alpha_s^2 \tilde{D}^{(1)}_0 + \alpha_e^2 \tilde{D}^{(2)}_0 + \alpha_A^2 \tilde{D}^{(3)}_0 + \dots}
\label{eq:dafbdmtt2} \eeq
for AxA model. The denominator now starts at ${\cal O}(\alpha^2)$, and the AxA model gives new ${\cal O}(\alpha_s \alpha_A)$-contributions to the numerator while ${\cal O}(\alpha_A^2)$-effects only modifies the total rate. The ${\cal O}(\alpha_s \alpha_X^2)$-contributions to the X-model asymmetry is again subleading. Each term in the numerator and denominator of \Eq{eq:dafbdmtt1} and \Eq{eq:dafbdmtt2} are denoted by $\Delta \sigma_{tot}$ and $\Delta \sigma_{FB}$ of QCD, QED and new physics, respectively.

Due to the leading-order calculation scheme, the only diagrams needed for the NLO X calculation are the ones depicted in \Fig{fig:nlo}. Those are leading contributors to the asymmetry. Meanwhile, other diagrams at the same coupling order give only subleading corrections to the total cross-section, hence ignored.

\begin{table}[t] \centering
\begin{tabular}{c|c| c| c}
\hline \hline
Model & Total $\Delta \sigma_{FB}$ & $0j$ excl.  & $1j+$ incl.  \\
\hline \hline
X+ & 55fb (7.9\%)&  97fb (11\%) & -42fb (-17\%)\\
X-- & -85fb (5.2\%)&  -183fb (6.3\%)& 98fb (-3.9\%)\\
\hline
AxA+ & 177fb (10\%) & 177fb (13\%) &  40fb (-9.4\%) \\
AxA-- & -92fb (5.5\%) & -92fb (7.9\%) & -12fb (-14\%) \\
\hline
QCD & 393fb (7.1$^{\pm 0.7}$\%) & 533fb (9.6$^{\pm 1.0}$\%)& -140fb (-13$^{- 0.9}$\%)\\
QED & 76fb (8.6\%)& 116fb (11.5\%)& -40fb (-17\%)\\
\hline \hline
\end{tabular}
\caption{Total inclusive asymmetry cross-sections  are divided into $0j$ exclusive and $1j+$ inclusive contributions where the division is conveniently set by the existence of a hard jet with $p_T(j)>20$GeV. Each model contributions are shown in each row, but resulting final asymmetries summing all contributions are also shown in parenthesis. For axigluons, no subleading higher-order corrections are added to the $0j$ results, and $\Delta \sigma_{FB}(tot) = \Delta \sigma_{FB}(0j)$. QCD scale uncertainties shown are about 10\% relatively~\cite{Brodsky:2012ik}. }
\label{tab:inc+exc} \end{table}
\begin{figure*}[t]
\includegraphics[width=0.9\columnwidth]{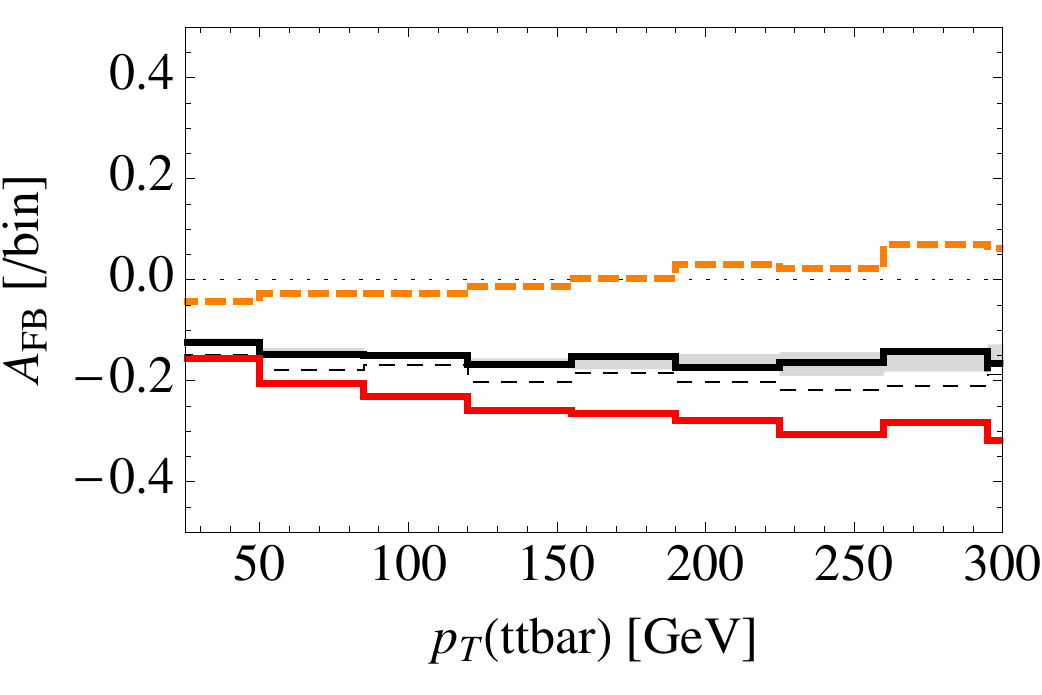}
\includegraphics[width=0.9\columnwidth]{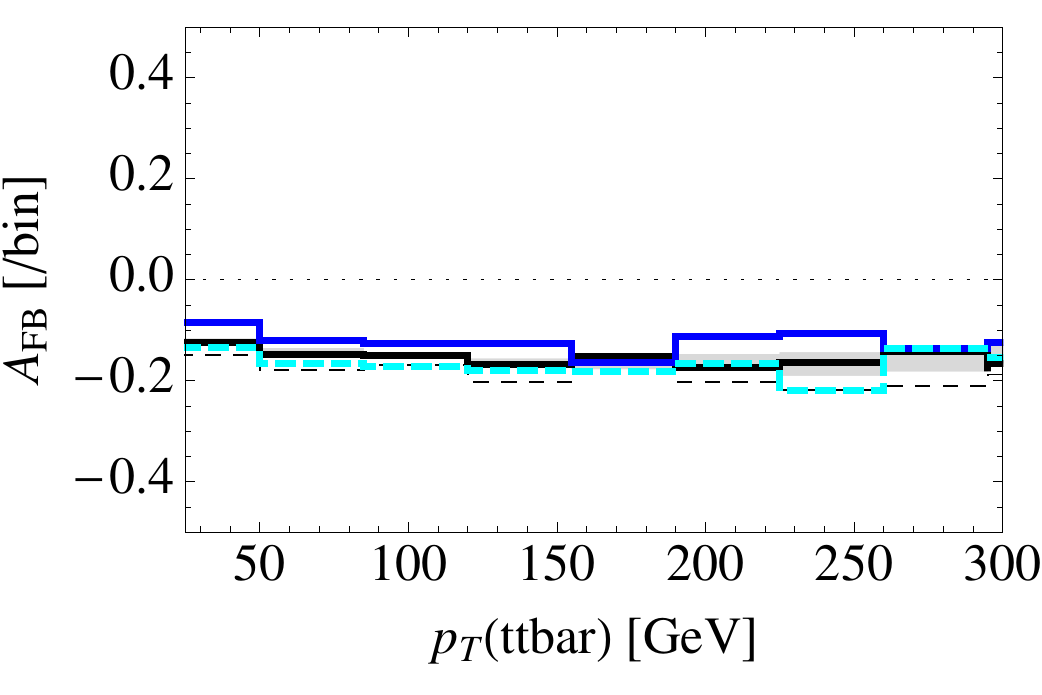}
\caption{Top asymmetry with top pair $p_T$ in the $1j+$ sample. The X(left) and AxA(right) models are compared with QCD(black solid) and QCD+QED(black dashed). Red, orange-dashed, blue and light-blue-dashed lines are X+, X--, AxA+ and AxA--. QCD contributions are added in all lines. The horizontal dotted reference line is at zero. Gray regions around QCD are scale uncertainties.}
\label{fig:dafbdptt} \end{figure*}

We tabulate numerical results of top asymmetries in Table~\ref{tab:benchmark} and Table~\ref{tab:inc+exc}. We show results with two specific choices of parameters for each models. All results are consistent with experimental data. Note also that we do not include known NLO corrections to QCD total cross-section according to our calculation scheme.

We discuss a notable feature shown in Table~\ref{tab:benchmark}. Both heavy X and AxA models need $\eta_t = -1$ to induce a positive asymmetry. It is well-known for AxA models~\cite{oai:arXiv.org:0911.2955}. The X model result can be contrasted with the QED's prediction of a positive asymmetry with $\eta_t= +1$. The only sign difference between X and QED comes from the propagator of 3TeV-X boson at 1.96TeV collision. The heavy propagator flips the sign of QCD-box and X-tree interference. This observation also implies  that if the X is much lighter than the top pair threshold, a positive $\AFB$ could be generated in a flavor-independent setup, $\eta_t=+1$, similarly to the QED. We will present detailed study of such X model in our future publication~\cite{Jung-RG}.

The full NLO results reported in Table~\ref{tab:benchmark} also support the observation of Ref.~\cite{Jung-RG} that leading log terms are likely subdominant and the renormalization group analysis of the operator mixing alone would wrongly predict a positive asymmetry from the heavy X model with $\eta_t=+1$.

\section{Correlation of asymmetries}

Table~\ref{tab:inc+exc} is the first place to glimpse the correlation we will discuss. As explained, the total cross-section is a sum of $0j$ exclusive and $1j+$ inclusive cross-sections (however, asymmetries are not simply added). But note that, according to our calculation scheme, for AxA model, $\Delta \sigma_{FB}$ starts at ${\cal O}(\alpha_s \alpha_A)$ for both the total and $0j$ exclusive, hence $\Delta \sigma_{FB}(tot) = \Delta \sigma_{FB}(0j)$. But the $1j+$ inclusive one starts at higher ${\cal O}(\alpha_s^2 \alpha_A)$ and is thus smaller. On the other hand, all three start at the same ${\cal O}(\alpha_s^2 \alpha_X)$ for X models and $\Delta \sigma_{FB}(tot) = \Delta \sigma_{FB}(0j) + \Delta \sigma_{FB}(1j+)$. 

The correlation is that, for X, QCD and QED, asymmetric cross-sections $\sigma_{FB}$ vary significantly among three samples and are similar in size as shown in Table~\ref{tab:inc+exc}. On the other hand, for AxA models, the $1j+$ $\sigma_{FB}$ is much smaller than that of the total and $0j$.

The correlation is more dramatically shown in \Fig{fig:dafbdptt} where the $1j+$ asymmetry distributions with $p_T(t\bar{t})$ are drawn. The 20GeV cut is reasonable to avoid the Sudakov region of small $p_T$~\cite{Skands:2012mm,Zhu:2012ts}. Although the X's total asymmetry is smaller than AxA's (see Table~\ref{tab:inc+exc}), its $1j+$ asymmetry can be much more enhanced compared to that of AxA as clearly depicted in the figure. Moreover, the X's contribution in this sample is much larger than the QCD scale uncertainties; thus can be measurable if experimental errors are well under control.

\begin{figure*}[t]
\includegraphics[width=0.9\columnwidth]{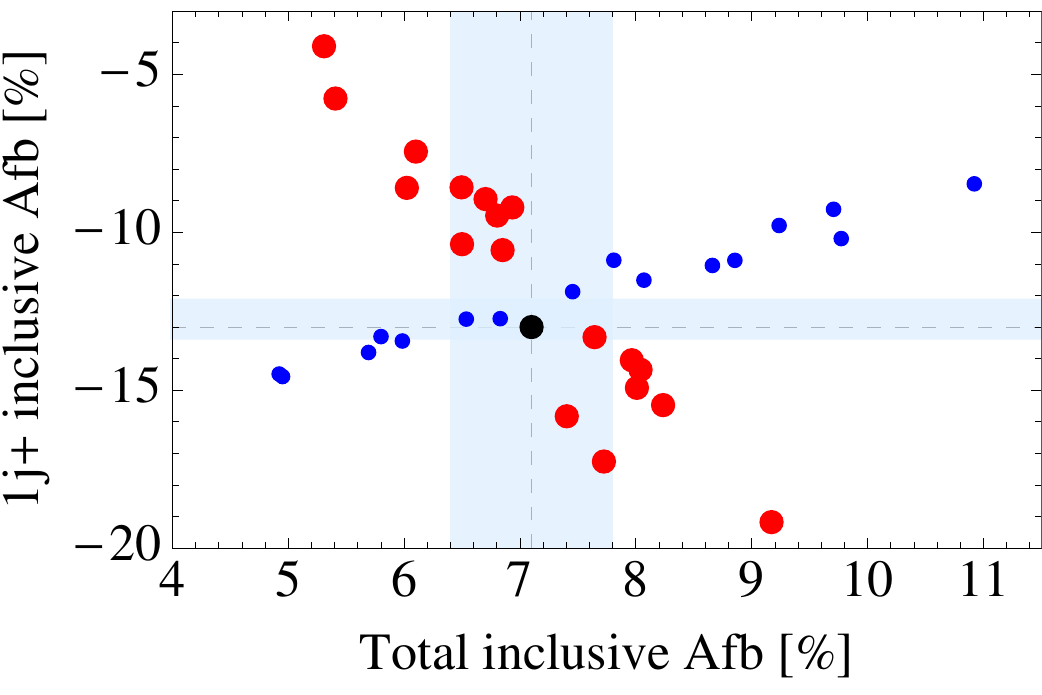}
\caption{Correlation of total inclusive and $1j+$ inclusive asymmetries. Correlations are clear and different between two models. Blue bands around the SM prediction (black) in the center are scale uncertainties. Shown models are X(red) and AxA(blue).}
\label{fig:correlation} \end{figure*}

Finally, we show \Fig{fig:correlation} where one can see that the correlation clearly exists and is different between the X and AxA models. The figure is drawn by randomly scanning model parameters within $1.5\leq M_X \leq 3.5\TeV$, $0.05 \leq \alpha_X \leq 0.5$ for X and $2.5 \leq M_A \leq 5\TeV$, $0.05 \leq \alpha_A \leq 0.5$ for AxA. $\eta_t = \pm1$ is also randomly selected. Both the sign and size of the correlation slope in \Fig{fig:correlation} are notably different. 

First of all, why the size is different can be understood more easily. For tree-models, the $1j+$ asymmetry is a subleading higher-order correction at ${\cal O}(\alpha^3)$ to the total asymmetry at ${\cal O}(\alpha^2)$. On the other hand, for loop-models, the $1j+$ asymmetry is one of two main contributions to the total asymmetry at the same ${\cal O}(\alpha^3)$; thus the $1j+$ and the total asymmetry are more comparable in size and the correlation slope in \Fig{fig:correlation} is steeper for loop-models. These can also be seen in \Eq{eq:dafbdptt}, \Eq{eq:dafbdmtt1} and \Eq{eq:dafbdmtt2}.

The sign of the correlation is deeply rooted under the structure of QCD singularities and color factors. It is useful to consider the soft singular limit of QCD radiation. In the soft limit, $q \to 0$, the squared gluon emission amplitude is factorized into the squared born amplitude and squared eikonal current\footnote{\Eq{eq:eikonallimit} and others below involving $\mathbf{T}_i$ are schematic. Color factors are not really factorized between eikonals and amplitudes, and we later put a superscript ${\cal M}^0_{born}$ when all color factors are explicitly calculated out as in \Eq{eq:m132}.}
  \beq 
  | {\cal M}_{real}|^2 \, \propto \, - \alpha_s  \mathbf{J}^\dagger(q) \cdot \mathbf{J}(q) \, | {\cal M}_{born}|^2,
  \label{eq:eikonallimit} \eeq
  where the eikonal current is (ignoring the top mass, for simplicity)
  \beq
  \mathbf{J}(q)^\mu \= \sum_{i} \mathbf{J}_{i}(q)^\mu \= \sum_i \mathbf{T}_i \, \frac{ p_i^\mu }{ p_i \cdot q },
  \eeq
  \bea
   \mathbf{J}^\dagger(q) \cdot \mathbf{J}(q) &=& \sum_{i,k} \mathbf{J}^\dagger_i(q) \cdot \mathbf{J}_k(q) \nonumber\\
   &=& \sum_{i,k} \mathbf{T}_i \cdot \mathbf{T}_k \, \frac{ p_i \cdot p_k}{ ( p_i \cdot q ) ( p_k \cdot q ) }.
   \eea
Thus, the $1j+$ asymmetry, produced by real corrections, is related to the born-level asymmetry by dipole colors, $\mathbf{T}_i \cdot \mathbf{T}_k$, and dipole kinematics, $W_{ik} \equiv \frac{p_i \cdot p_k}{(p_i \cdot q)(p_k \cdot q)}$. It is clear that the energy of a gluon, $\omega$, does not change the sign of asymmetry. By integrating over the direction of gluons~\cite{Skands:2012mm}
\beq
F_{ik} \equiv \int W_{ik} \, d \Omega \, \simeq \, \frac{8\pi}{\omega^2} \left( \log \left( \frac{2 p_i \cdot p_k}{ m_i m_k } \right)  -1 \right),
\eeq
where we keep the masses of quarks to regularize IR divergences, we have the simple dependence on $p_i \cdot p_k$ which is relevant to the asymmetry.

If the born process does not generate any asymmetry as in QCD, the $1j+$ asymmetry should be generated from eikonals asymmetric under $t \leftrightarrow \bar{t}$. Such asymmetric eikonals are $\{ i,k \} = (1,3), (1,4), (2,3), (2,4)$, where 1,2,3,4 denote $q,\bar{q},t,\bar{t}$ in the $q\bar{q} \to t \bar{t}$. When the $F_{ik}$ is integrated over the forward-backward(FB)-asymmetric phase space, we have
\beq
\left[ \int_0^1 - \int_{-1}^0 \right] \, F_{ik} \, d \cos \theta_t \left\{ \bmat  >0 & {\rm for}\, \{i,k\} = (1,4), (2,3) \\  <0 & {\rm for}\, \{i,k\} = (1,3), (2,4) \\ =0 & {\rm otherwise}  \emat \right. .
\eeq
For example, the interference between ${\cal M}_1$ and ${\cal M}_3$ (where the subscripts imply the parton emitting a gluon), denoted by $M_{13}$,  is approximately (in the soft limit)
  \bea M_{13} &\propto& -\alpha_s \mathbf{T}_1 \cdot \mathbf{T}_3 \, F_{13}(t) \, | {\cal M}_{born}|^2  \nonumber \\
  &=& + \frac{1}{16}( f_{abc}^2 + d_{abc}^2 ) \, \alpha_s \, F_{13}(t) \, | {\cal M}^0_{born} |^2 \nonumber\\
  &=& + \frac{7}{3} \alpha_s \, F_{13}(t) \, | {\cal M}^0_{born} |^2 \label{eq:m132},
  \eea
  and similarly
  \bea M_{14} &\propto& -\alpha_s \mathbf{T}_1 \cdot \mathbf{T}_4 \, F_{14}(u) \, | {\cal M}_{born}|^2 \nonumber \\
 &=& - \frac{1}{16}( -f_{abc}^2 + d_{abc}^2 ) \, \alpha_s \, F_{14}(u) \, | {\cal M}^0_{born} |^2 \nonumber\\  
 &=& + \frac{2}{3} \alpha_s \, F_{14}(u) \, |{\cal M}^0_{born}|^2 \label{eq:m142}.
  \eea
Since the born process is FB-symmetric, they add to generate non-zero and negative $1j+$ asymmetry as is well known. The same calculation holds for $\{i,k\} = (2,3),(2,4)$. All other pairs of $\{i,k\}$ give symmetric eikonals, hence no asymmetry. Combined with a positive total asymmetry, a negative correlation slope is derived.

In this argument, the eikonal approximation \Eq{eq:eikonallimit} and the dipole kinematics, $F_{ik}$, are solely dictated by QCD. The model dependencies reside in dipole color factors and squared born amplitudes. For the X model, dipole colors are $\mathbf{T}_1 \cdot \mathbf{T}_3 = -2$ and $\mathbf{T}_1 \cdot \mathbf{T}_4 = +2$, and we have
  \bea M_{13} &\propto& -\alpha_s \mathbf{T}_1 \cdot \mathbf{T}_3 \, F_{13}(t) \, | {\cal M}_{born}|^2  \nonumber \\
  &=& + 2 \alpha_s \, F_{13}(t) \, | {\cal M}^0_{born} |^2,
  \eea
  \bea M_{14} &\propto& -\alpha_s \mathbf{T}_1 \cdot \mathbf{T}_4 \, F_{14}(u) \, | {\cal M}_{born}|^2 \nonumber \\
 &=& -2 \alpha_s \, F_{14}(u) \, |{\cal M}^0_{born}|^2. 
  \eea
The $\eta_t=-1$ cancels with the minus sign from a heavy propagator in the interference amplitudes. Therefore, a negative $1j+$ asymmetry is again generated, and all other sign arguments follow that of the QCD above; thus, a negative correlation slope is derived as shown in \Fig{fig:correlation}.

What about the axigluon AxA? For axigluon, the limiting expressions in \Eq{eq:eikonallimit}, \Eq{eq:m132} and \Eq{eq:m142} are again the same as dictated by QCD. However, the born process now already generates an asymmetry and $|{\cal M}^0_{born}|^2 \, \propto \pm \beta_t c_t \equiv \pm \sqrt{1- 4m_t^2/\hat{s}} \cos \theta_t$ with $\eta_t = \mp 1$ at the QCD-AxA interference level. The limiting expressions can further be written as
  \bea
  M_{13} &\propto& \pm \frac{7}{3} \alpha_s F_{13}(t) \, \beta_t c_t, \\
  M_{14} &\propto& \pm \frac{2}{3} \alpha_s F_{14}(u) \, \beta_t c_t.
  \eea
 Additional $\beta_t c_t$ factor makes them to generate the top asymmetry with the same sign as the born-level asymmetry, i.e., $\pm$. Thus, a positive correlation slope is derived for AxA as shown in \Fig{fig:correlation}.

Unlike in the case of QCD, there are more contributions for AxA. When symmetric eikonals are multiplied by the asymmetric born process, non-zero top asymmetry is also induced. One example symmetric eikonals with $\{i,k\} = (1,2)$ is approximated as 
  \bea M_{12} &\propto& -\alpha_s \mathbf{T}_1 \cdot \mathbf{T}_2 \, F_{12}(\hat{s}) \, | {\cal M}_{born}|^2 \nonumber\\
  &\propto& - \frac{1}{3} \alpha_s F_{12}(\hat{s}) \, | {\cal M}^0_{born}|^2 \, \propto \, \mp \frac{1}{3} \alpha_s F_{12}(\hat{s}) \, \beta_t c_t,
  \eea
  which is FB-asymmetric with the opposite sign. But this negative coefficient is smaller than previous coefficients; thus, they do not change the sign of final $1j+$ asymmetry\footnote{For the given partonic collision energy $\hat{s}$, the integrated functions $\int F_{ik} \, \beta_t c_t \, d c_t$ are positive and similar in size for all $(i,k)$.}. Other symmetric eikonals with less singular $g \to t\bar{t}$ splitting are less influential.

It was useful to consider a soft singular limit because the eikonal approximation in \Eq{eq:eikonallimit} directly relates the born process with the $1j+$ process. Does the prediction in the soft limit persist to any three-body phase space? As far as we are concerned with the sign of the asymmetry, it is likely so at least for the majority of phase space nearby the soft limit which is a dominant contributor to the top asymmetry in QCD~\cite{Skands:2012mm,Melnikov:2010iu,Li:2013mia}. We thus assume that the soft region can be usefully used in our argument.

The correlation can be generalized to any tree-models and to the most important class of loop-models. Any tree models will have $|{\cal M}^0_{born}|^2$ piece whose dominant terms are proportional to $\pm \beta_t c_t$; then the same argument used for AxA above will apply (regardless of how the tree-level asymmetry is generated). On the other hand, our discussion for the loop-model using the $VV^{(1)}$ model is already quite general. It is because the $VV^{(1)}$ is found to be the only interesting loop-model in Ref.~\cite{Jung-RG}. Although $AA^{(1)}$ can, in principle, also generate the asymmetry at higher order, the size of induced asymmetry is too small to play an interesting role\footnote{This can also be seen from the eikonal approximation. $M_{13} \propto \pm 2 \alpha_s F_{13}(t) \, \beta_t c_t$ and $M_{14} \propto \mp 2 \alpha_s F_{14}(u) \, \beta_t c_t$ add to cancel any asymmetries at this order, whereas $M_{12}=0$.}~\cite{Jung-RG}. Thus, our previous discussion applies generally to any interesting loop and tree models.

We also briefly comment that the correlation sign can be flipped in some non-standard models where some high color representation yields dipole color factors with different signs.

The AxA and the X models considered are extremum cases of the tree- and loop-models for our study. If some tree-model has both $AA^{(8)}$ and $VV$ interactions such as AxR model, the correlation may then extrapolate between that of AxA and X with the relative strengths of $AA^{(8)}$ and $VV$ interactions.

It has been recently discussed that ${\cal O}(\alpha_s^4)$ corrections to the $t\bar{t} + j$ has sizable impacts on the top asymmetry~\cite{Dittmaier:2007wz,Dittmaier:2008uj,Melnikov:2010iu}. The NLO(${\cal O}(\alpha_s^4)$) corrections almost cancel the LO(${\cal O}(\alpha_s^3)$) asymmetry in the $t\bar{t}+j$. The worry was that the lowest-order asymmetry might then be generically unstable. Later, however, it was argued that the cancellation is accidental and will not persist to even higher-orders, and that the LO inclusive asymmetry is stable~\cite{Melnikov:2010iu}. If such cancellation also happens for others models, the advocated correlation and spectrum may have reduced sensitivities. Although the cancellation in QCD is likely accidental and does not likely persist to new physics models, a definite answer can be obtained only by thorough calculations.

\begin{figure*}[t]
\includegraphics[width=0.9\columnwidth]{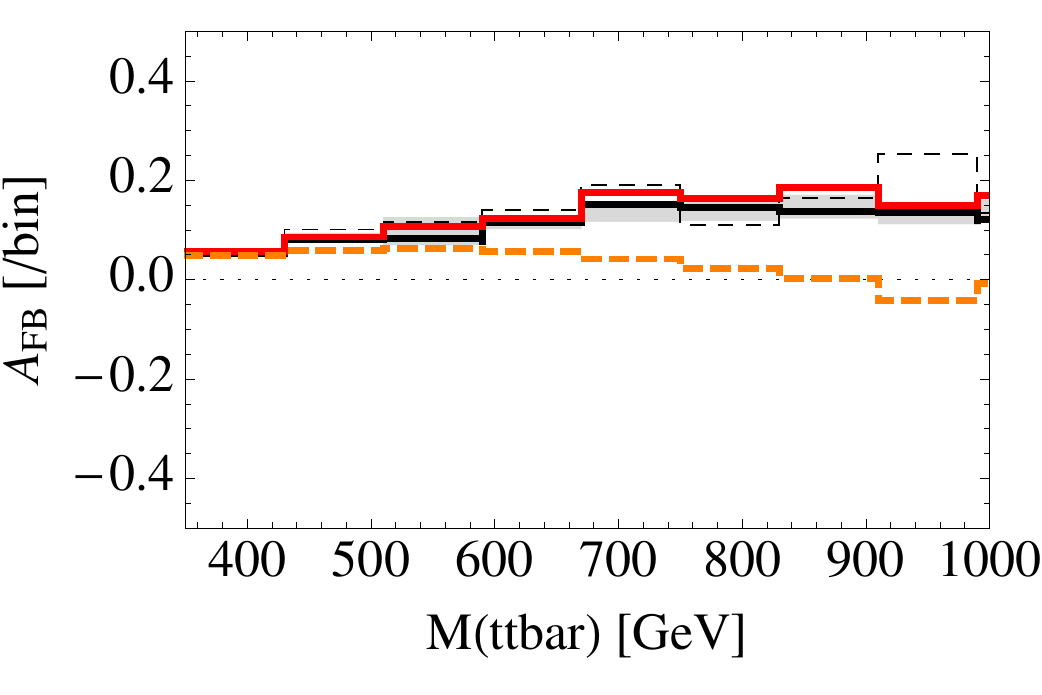}
\includegraphics[width=0.9\columnwidth]{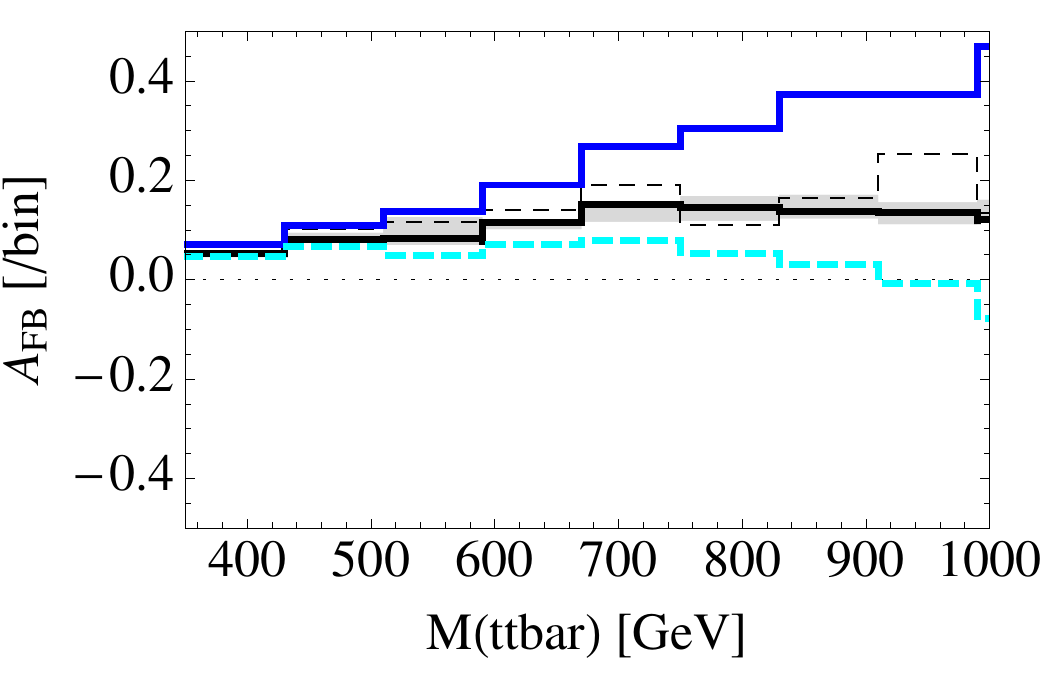}
\caption{Same as \Fig{fig:dafbdptt}, but top asymmetry with total top pair invariant mass.}
\label{fig:dafbmtt} \end{figure*}

\begin{figure*}[t]
\includegraphics[width=0.9\columnwidth]{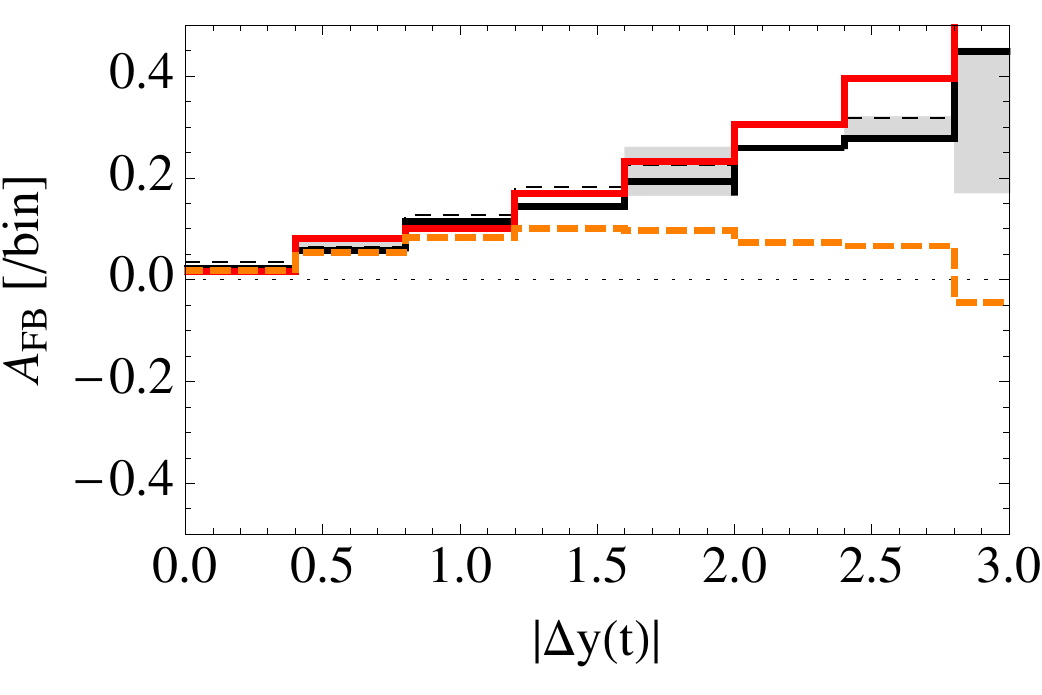}
\includegraphics[width=0.9\columnwidth]{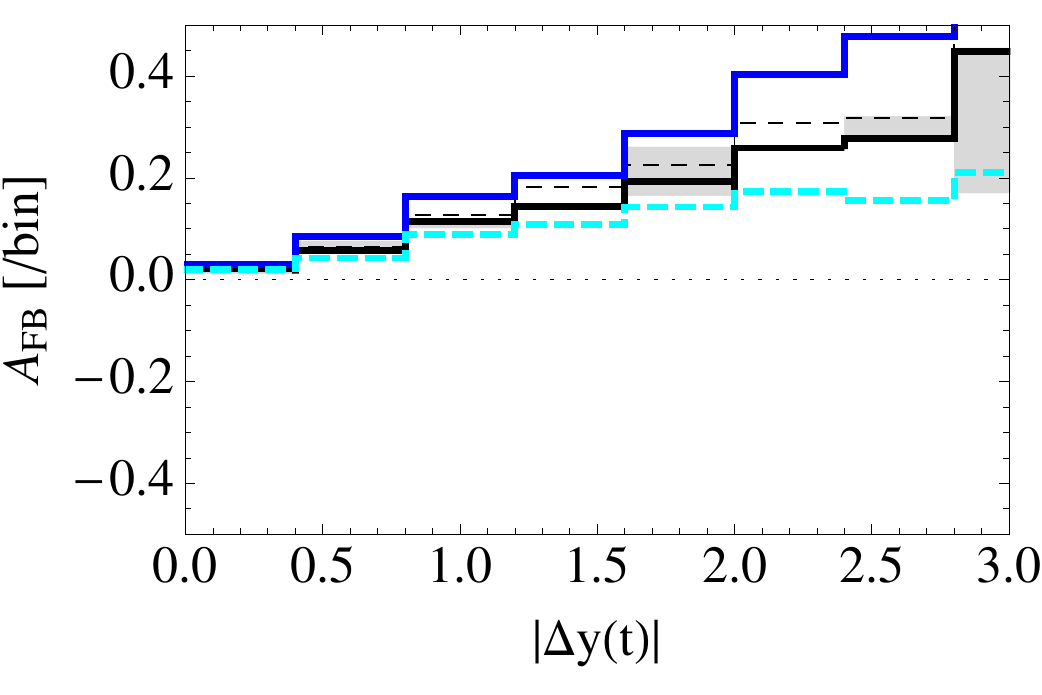}
\caption{Same as \Fig{fig:dafbdptt}, but top asymmetry with total top pair rapidity difference.}
\label{fig:dafbytt} \end{figure*}

\begin{figure*}[t]
\includegraphics[width=0.9\columnwidth]{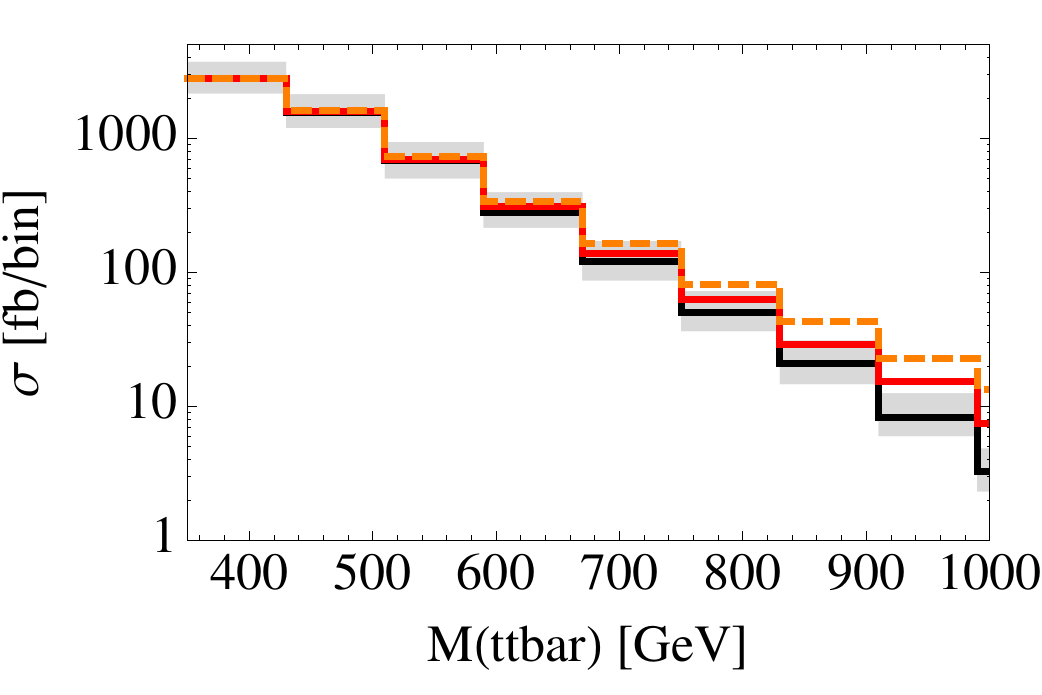}
\includegraphics[width=0.9\columnwidth]{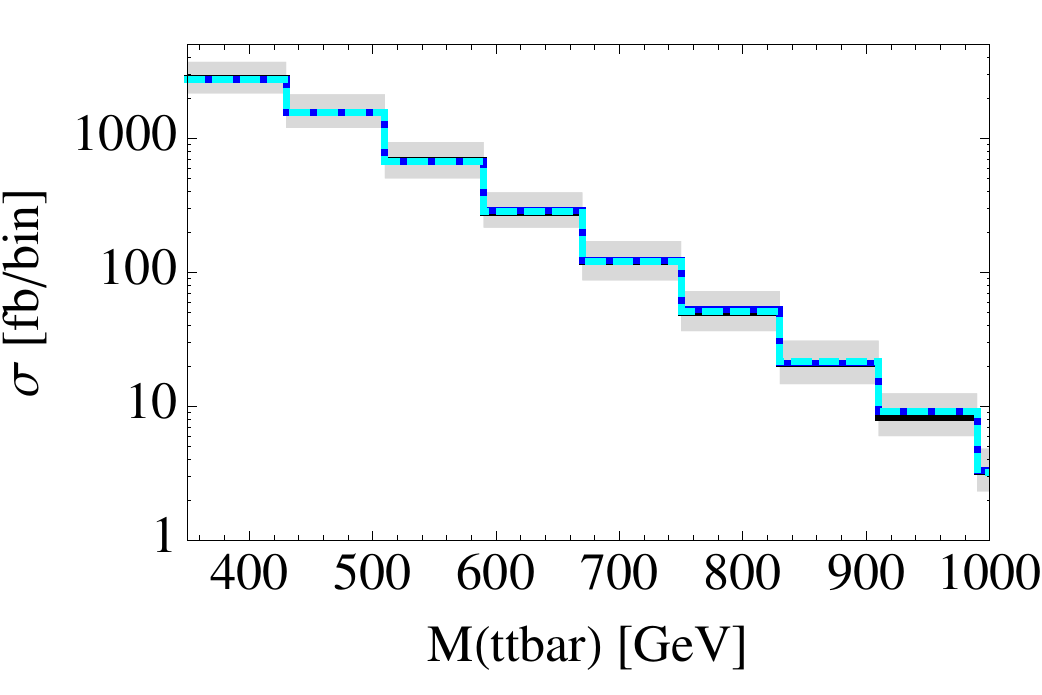}
\caption{Same as \Fig{fig:dafbdptt}, but cross-section with total top pair invariant mass.}
\label{fig:dmtt} \end{figure*}

\begin{figure*}[t]
\includegraphics[width=0.9\columnwidth]{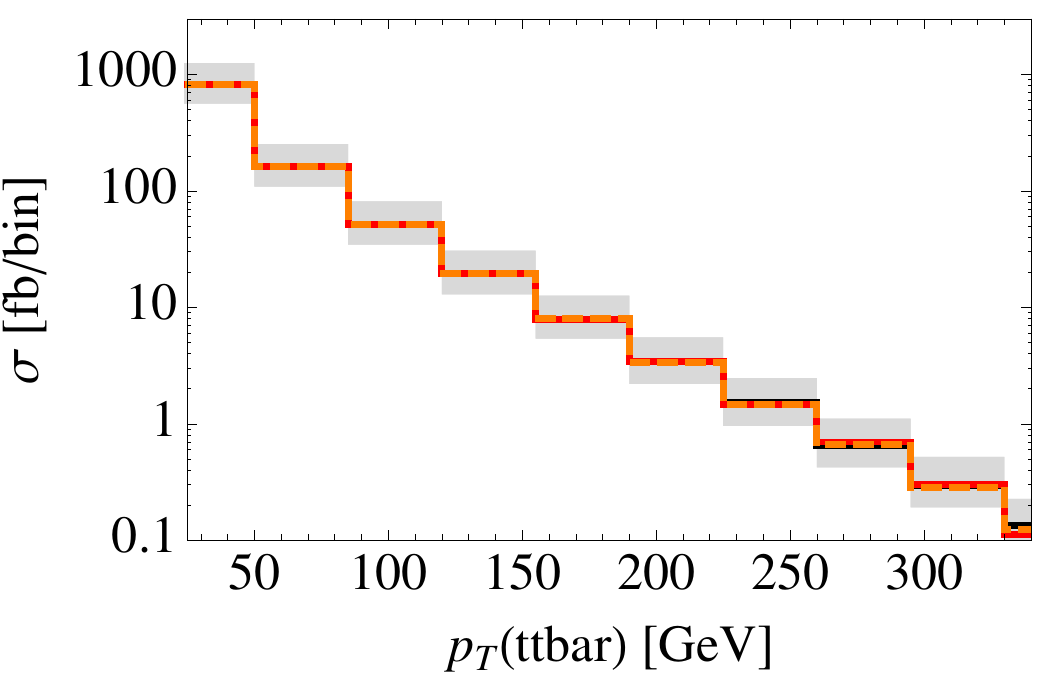}
\includegraphics[width=0.9\columnwidth]{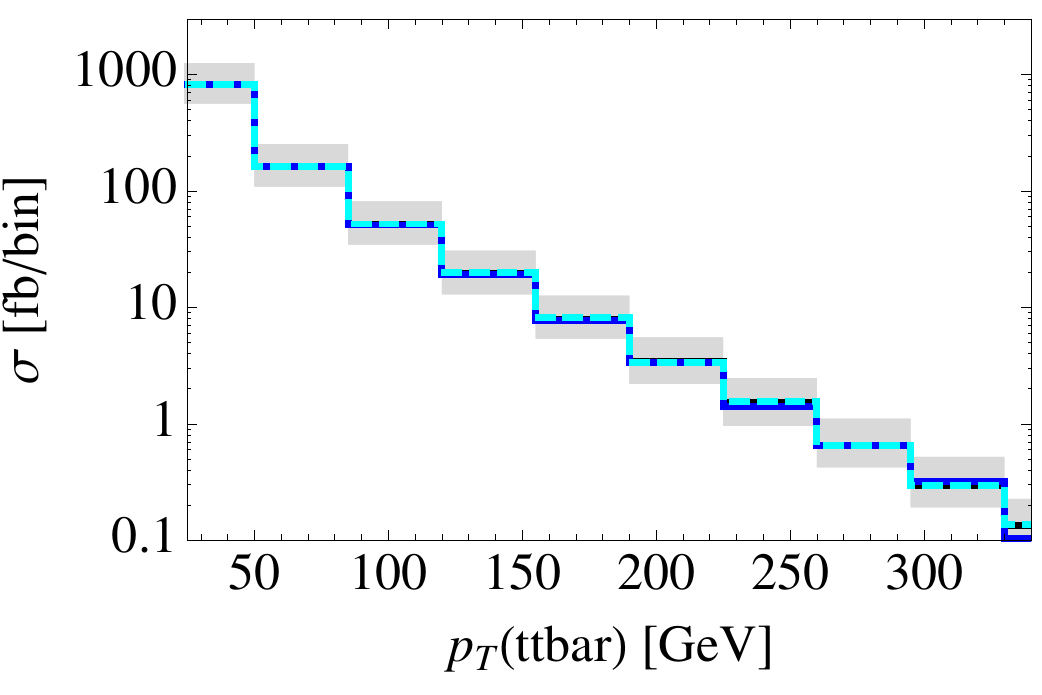}
\caption{Same as \Fig{fig:dafbdptt}, but cross-section with top pair $p_T$ in the $1j+$ sample.}
\label{fig:dptt} \end{figure*}

It is still true that there are other conventional spectra that can tell the existence of new asymmetry contributions. For example, $d A_{FB} / d m(t\bar{t})$ and  $d A_{FB} / d |\Delta y(t)| $ which are being measured at Tevatron can show deviations due to new physics models as depicted in \Fig{fig:dafbmtt} and \Fig{fig:dafbytt}. We, however, emphasize that they do not efficiently distinguish the tree vs. loop origins of new asymmetries. Even more difficultly, differential total rate spectra such as $d\sigma / d m(t\bar{t})$ and $d \sigma / d p_T(t\bar{t})$ shown in \Fig{fig:dmtt} and \Fig{fig:dptt} may not even see clear evidences of such new physics or will not have clear connections with top asymmetry.


\section{LHC14 prospects}

\begin{table}[t] \centering
\begin{tabular}{c|c| c| c}
\hline \hline
Model & Total $A_C^{\Delta |y|}$ & $1j+$ incl. \\
\hline \hline
$X+$ &  0.86\% & -0.66\% &\\
\hline
AxA+ &  0.41\% & -0.29\% & \\
\hline
QCD &  0.62\% & -0.27\% & \\
\hline \hline
\end{tabular}
\caption{At LHC14. Although overall asymmetries are smaller than Tevatron ones, the correlation between total and $1j+$ asymmetries persists.}
\label{tab:LHC14} \end{table}

We calculate the following asymmetry observable at LHC14.
\beq 
A_C^{\Delta |y|} = \frac{ N( \Delta |y| >0 ) - N( \Delta |y| < 0 ) }{ N( \Delta |y| >0 ) + N( \Delta |y| < 0 ) },
\eeq
where $\Delta |y| \equiv |y(t)| - |y(\bar{t})|$. This observable has been used to measure charge asymmetry at LHC~\cite{Aad:2013cea,CMS:2013nfa}. Various other observables that can be correlated with QCD charge asymmetry have been considered~\cite{Bernreuther:2012sx,Antunano:2007da}, but similar conclusion made with $A_C^{\Delta |y|}$ will be applied to them.

See Table~\ref{tab:LHC14}. At LHC14, the sizes of asymmetries are smaller. But the correlation of total and $1j+$ inclusive  asymmetries persists and differs between tree- and loop-models considered; although X+ and AxA+ generate similar total $A_C^{\Delta |y|}$ (just with opposite sign), the X+ produces a much higher asymmetry in the $1j+$ sample. Proper cuts enhancing the top asymmetry measurements at LHC will also help to measure the correlation.

\section{Conclusions and discussions}

We have studied how the correlation of total and $1j+$ asymmetries can reveal the origin of asymmetries whether as tree- or loop-induced. We considered leptophobic $Z^\prime$, denoted by X, and pure axigluon, denoted by AxA, as benchmark models for loop- and tree-induced asymmetries. By comparison study, we found that both the sign and size of the correlation slope are clearly different between two models as nicely depicted in \Fig{fig:correlation}. The correlation was understood in the QCD eikonal approximation which directly relates the $t\bar{t}j$ and $t\bar{t}$ processes. The discussion in the eikonal limit was generalized to any tree-level $A_{\rm FB}$ models and to the most important class of loop-level models; in general, the correlation exists and is different between loop- and tree-models.

If the correlation is proven useful, there are several ways to improve the study. We have included only leading contributions to all observables. The sensitivities of the advocated correlation and spectra, however, may depend on yet unknown even higher-order corrections. More dedicated calculation and model consideration will be useful. The discrimination feature of the correlation will also further be improved upon by including various other spectra and channels to build more sophisticated correlators.

The study also implies that the loop-induced asymmetry may be better measurable in the inclusive $t\bar{t}j$ channel than in the inclusive $t\bar{t}$ channel. \Fig{fig:correlation} and Table~\ref{tab:inc+exc} show that the loop-asymmetry stands more clearly above QCD scale uncertainties in the $t\bar{t}j$ inclusive channel. Although more realistic collider analysis should be carried out for a better estimation, it is useful to know this possibility. The correlation and the better measurability discussed at Tevatron apply to LHC14 as well; thus, future dedicated measurements at LHC are encouraged. 

A necessary condition for the positive total asymmetry from loop-models is also discussed based on our full NLO calculation of the X model: $\eta_t =-1$ for the heavy X. The full NLO result was compared and contrasted with the prediction based on renormalization group operator mixing made in Ref.~\cite{Jung-RG}, and more consistent effective theory calculation was motivated.

The leptophobic $Z^\prime$ models may first be discovered through dijet or top pair resonance searches at hadron collider. We, however, emphasize that such total rate measurements do not tell us whether the model is responsible for the top asymmetry and whether the asymmetry is loop- or tree-level induced. In any case, top asymmetries and their correlations will provide unique and valuable information of $Z^\prime$ coupling structure.

Model building options for loop-$A_{\rm FB}$ are different from tree-$A_{\rm FB}$ model building options. The measurability of loop-induced asymmetries seems higher than usually expected. No compelling reason and no satisfactory possibility of large tree-level asymmetries are present. The phenomenological study of loop-models may thus be more seriously pursued. Our study hopefully brings a useful step towards it.

%

\begin{acknowledgments}
\emph{Acknowledgements}
We thank Chul Kim and Hua Xing Zhu for useful discussions and Giulia Zanderighi for introduction of necessary tools. SJ also acknowledges James Wells for encouraging the study of related subjects. SJ and YWY thank KIAS Center for Advanced Computation for providing computing resources. 

\vspace{1mm}
SJ is supported in part by National Research Foundation (NRF) of Korea under grant 2013R1A1A2058449. PK is supported in part by Basic Science Research Program through the NRF under grant 2012R1A2A1A01006053 and by SRC program of NRF funded by MEST (20120001176) through Korea Neutrino Research Center at Seoul National University. CY is supported by Basic Science Research Program through the NRF funded by the Ministry of Education Science and Technology 2011-0022996. 

\end{acknowledgments}


\end{document}